\documentclass{IEEEtran}
\usepackage{cite}
\usepackage{amsmath,amssymb,amsfonts}
\usepackage{graphicx}
\usepackage{textcomp,nicefrac}
\usepackage{url}
\newcommand{\muec}{$\mu\rightarrow e$ conversion}
\def\BibTeX{{\rm B\kern-.05em{\sc i\kern-.025em b}\kern-.08em
T\kern-.1667em\lower.7ex\hbox{E}\kern-.125emX}}
\begin{document}
\title{An FPGA-based Trigger System with Online Track Recognition in COMET Phase-I}
\author{Yu Nakazawa, Yuki Fujii, Masahiro Ikeno, Yoshitaka Kuno, MyeongJae Lee, Satoshi Mihara, Masayoshi Shoji, Tomohisa Uchida, Kazuki Ueno, Hisataka Yoshida
\thanks{This work was submitted on May 26, 2021.}
\thanks{This work was supported by JSPS KAKENHI Grant Numbers JP17H04841, JP17H06135, JP18H05231, JP18J10962.}
\thanks{Y. Nakazawa is with the Department of Physics, Osaka University, Osaka, 560-0043, Japan (e-mail: y-nakazawa@epp.phys.sci.osaka-u.ac.jp).}
\thanks{Y. Fujii is with the School of Physics and Astronomy, Monash University, Clayton, Victoria, 3800, Australia (e-mail: Yuki.Fujii@monash.edu).}
\thanks{M. Ikeno, S. Mihara, M. Shoji, T. Uchida, and K. Ueno are with Institute for Particle and Nuclear Studies, High Energy Accelerator Research Organization (KEK), Tsukuba, Ibaraki, 305-0801, Japan.}
\thanks{Y. Kuno and H. Yoshida are with Research Center for Nuclear Physics, Osaka University, Osaka, 567-0047, Japan.}
\thanks{M.J. Lee is with Center for Axion and Precision Physics Research (CAPP), Institute for Basic Science (IBS), Daejeon, 34126, South Korea.}
}

\maketitle

\begin{abstract}
An FPGA-based online trigger system has been developed for the COMET Phase-I experiment, which searches for muon-to-electron conversion as a signature of New Physics beyond the Standard Model. 
A drift chamber and trigger counters are devised to detect a mono-energetic electron from the conversion process in a 1-T solenoidal magnetic field.
A highly intense muon source enables reaching unprecedented experimental sensitivity, however it also generates undesirable background particles which increase the trigger rate to a much higher level than the capability of the data acquisition system.
By using hit information from the drift chamber in addition to the trigger counter, the newly proposed online trigger system efficiently suppresses the background trigger rate while keeping the signal-event acceptance large.
A characteristic of this system is the utilization of machine learning techniques for optimizing lookup tables implemented in hardware.
Simulation studies show that the signal-event acceptance of the online trigger is 96\% while the background trigger rate is reduced from 91\,kHz to 13\,kHz.
The global trigger system and trigger electronics that construct a distributed trigger architecture were built.
The total trigger latency was estimated to be $3.2\,\mathrm{\mu s}$. 
The trigger system test was carried out by using a part of the drift-chamber readout region.
\end{abstract}


\section{Introduction} \label{sec:introduction}
\IEEEPARstart{T}{HE} COherent Muon To Electron Transition (COMET) Phase-I experiment~\cite{comet:tdr} plans to search for the neutrinoless transition of a muon into an electron (\muec) in an aluminum muonic atom.
In the Standard Model, the lepton flavor violating processes between generations of charged leptons are only allowed through the neutrino mixing due to its finite mass, while its expected branching ratio is extremely suppressed because of the tiny neutrino mass.
Therefore, the \muec\ process in the Standard Model can not be observed experimentally~\cite{phys:mu-e}.
On the other hand, many New Physics models beyond the Standard Model predict this process at a detectable rate.
Therefore, a discovery of this conversion would be a clear signature of New Physics beyond the Standard Model.
The most stringent limit of the \muec\ branching ratio, $7.0\times10^{-13}$ with a 90\% confidence level~\cite{sindrum-ii}, is given by the SINDRUM-II experiment using a gold target.
The goal of COMET Phase-I is to improve the sensitivity upper limit by a factor of 100, corresponding to $7.0\times10^{-15}$ (90\% confidence level), with a 150 day-long physics measurement by using the proton beam facility of Japan Proton Accelerator Research Complex (J-PARC) in Tokai, Japan.

\begin{figure}[t]
 \centering
 \includegraphics[width=0.4\textwidth]{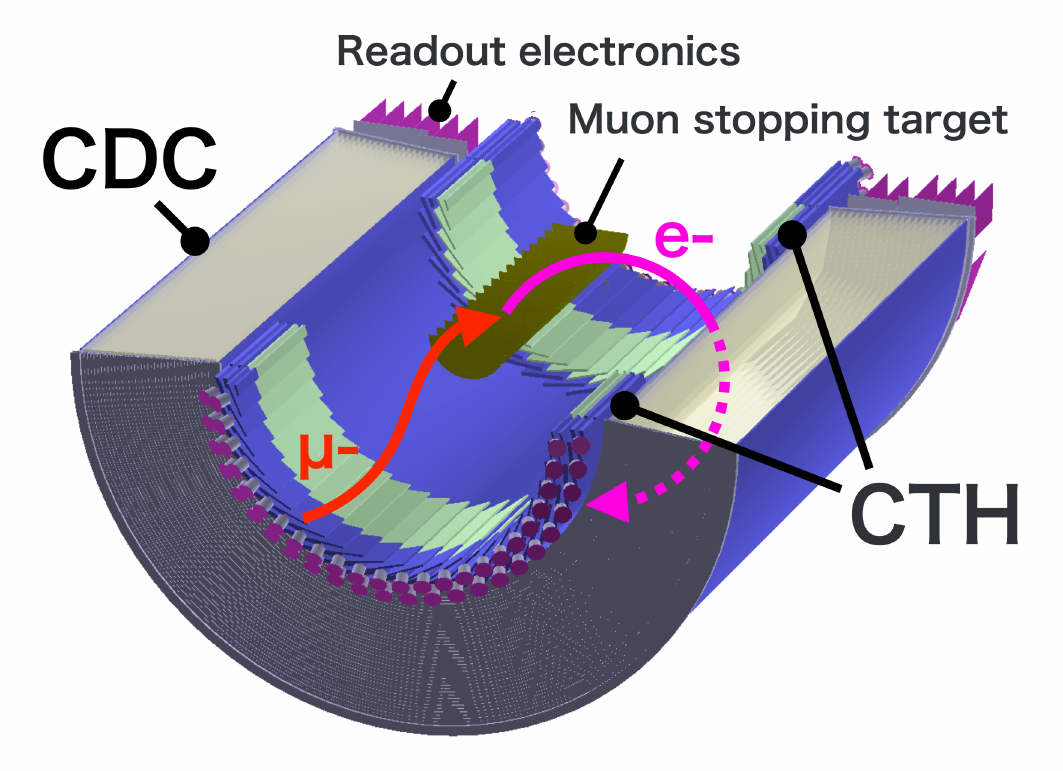}
 \caption[Schematic layout of the cylindrical detector system] {Schematic layout of the cylindrical detector system. The red and magenta arrows are respectively a muon track and a conversion-electron track for illustrative purposes.}
 \label{fig:cydet}
\end{figure}

The signal for \muec\ is a mono-energetic electron of 105\,MeV/c which is emitted from a muonic atom formed in an aluminum target.
This electron is detected by a cylindrical detector system (CyDet) sitting in a 1-T solenoidal magnetic field.
The CyDet consists of a cylindrical drift chamber (CDC) and a cylindrical trigger hodoscope (CTH), as shown in \figurename~\ref{fig:cydet}.
%
%
The CDC measures hit positions of charged particles with 4986 sense wires strung in a cylindrical structure of stereo wire planes and evaluates the momentum of the conversion electrons by tracking.
The RECBE board developed by the Belle II group~\cite{recbe} is adopted as the CDC readout electronics after some modifications on the digital logic~\cite{comet:tdr}.
The CTH is placed at the inner cylindrical wall of the CDC both upstream and downstream.
It provides the primary trigger signal and determines the precise timing of the charged particles.
The CTH has an inner scintillation layer and an outer Cherenkov layer.
The inner layer consists of 48 segmented plastic scintillators with a thickness of 5\,mm, and the outer layer consists of 48 segmented acrylic plastic Cherenkov radiators with a thickness of 10\,mm. 
Scintillation and Cherenkov radiation photons are read by fine-mesh photomultiplier tubes attached at the end of light guides.
Given the refractive index of an acrylic plastic (1.49), non-relativistic charged particles do not emit Cherenkov radiations.
By taking a four-fold coincidence (two sets of scintillator and Cherenkov rings) within 10\,ns, these counters trigger the conversion electron signal.

In order to reach the sensitivity goal, a highly intense pulsed muon beam will be generated from an 8\,GeV, 3.2\,kW pulsed proton beam provided by the J-PARC Main Ring.
The stopped muon rate in the aluminum target is expected to be more than $10^9$ per second.
These muons and remaining pions generate background particles: electrons, photons, and protons.
Such particles directly or indirectly induce a lot of undesired background hits in the detectors, resulting in an extremely high hit rate in any single CTH counter, around 2\,MHz.
Low-energy electrons from photon interactions with the detectors and their supporting structures are especially serious.
Many of these electrons deposit energy in CTH counters, resulting in passing the four-fold coincidence requirement accidentally, subsequently being misidentified as high-energy signal electrons.
According to a simulation study, a four-fold coincidence rate due to the background particles was estimated to be 91\,kHz.
The data acquisition (DAQ) system, however, requires a trigger rate of less than 26\,kHz so that conversion-electron events cannot be missed due to the DAQ dead time.

In order to solve this issue, we have developed an online trigger system called the COMET trigger (COTTRI) system, where the CDC-hit information is classified using machine learning (ML) techniques~\cite{phdthesis:ewen,trigger-ieee-proc,trigger-eps-hep-proc}.
The low-energy electrons which result in accidental trigger due to CTH hits hardly reach the CDC due to the magnetic field.
Such CTH-hit positions have less correlation to a hit pattern in the CDC, and therefore, the use of the CDC-hit information contributes to the trigger-rate reduction.
The maximum allowed trigger rate is set to be 13\,kHz including a safety factor of two.
The trigger latency of the full system is required to be less than 7$\,\mathrm{\mu s}$.
This value comes from a buffer size of 8.5$\,\mathrm{\mu s}$ on the RECBE board and a recorded event window of 1.1$\,\mathrm{\mu s}$.

The conceptual design was studied in Ref.~\cite{trigger-eps-hep-proc} and showed promising results.
Since then, the simulation study has been repeated with higher statistics to optimize the parameters and obtain more precise estimations.
In addition, we have developed a new board based on a field programmable gate array (FPGA) and tested the full chain performance.
Here, we show the hardware logic of the trigger system, the trigger algorithm of the COTTRI system, the expected system performance, and the trigger-related electronics.

\section{COMET Trigger System} \label{sec:cottri}
\begin{figure}[t]
 \centering
 \includegraphics[width=0.45\textwidth]{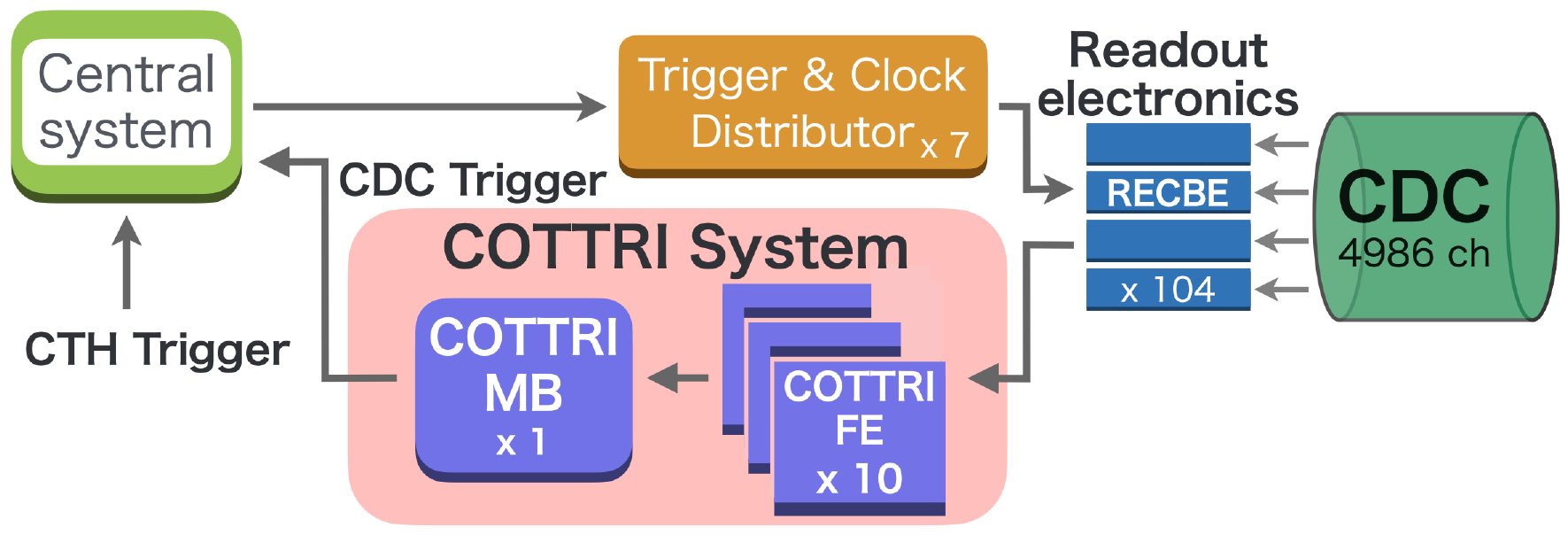}
 \caption[Conceptual design of the trigger system for the CyDet] {Conceptual design of the trigger system for the CyDet.}
 \label{fig:cydet-trigger-system}
\end{figure}

\figurename~\ref{fig:cydet-trigger-system} illustrates the conceptual design of the trigger system for the CyDet.
%
%
Initially, the RECBE boards receive the signals from the CDC, digitize the voltage into a 10-bit ADC sample at 30\,MHz, and store them internally for DAQ transaction.
In parallel, three samples are aggregated and compressed into 2\,bit words optimized to effectively separate the minimum ionization of signal-like electrons from electric noise and large energy depositions likely to be caused by protons or other heavier background particles.
Thus, the RECBE boards generate 2-bit energy-deposition information for each CDC wire and send it to the COTTRI system every 100\,ns.

The COTTRI system consists of ten COTTRI Front-End (FE) boards and a COTTRI Merger-board (MB). 
This distributed trigger architecture enables the system to cover the high number of CDC readout channels for making a trigger decision based on the CDC-hit information. 
In parallel, the CTH electronics provide the trigger signal with a much simpler scheme, namely, discriminating the analog signals from fine-mesh photomultiplier tubes, merging hit information over all channels, and taking a geometrically neighboring four-fold coincidence in a 10\,ns time window.
The central trigger system receives primitive trigger signals from the CDC and CTH sub-systems separately, and subsequently generates the final trigger decision by taking a coincidence of these primitive triggers.
The final trigger decision is distributed to all the readout electronics through distributor boards.
The trigger decisions and related processes are performed in the pipeline and are asynchronous with the proton bunch.
The central trigger system also distributes a 40\,MHz clock signal, thus all the processes are synchronized in the entire trigger system at this rate.

\section{Trigger algorithm} \label{sec:algo}
\begin{figure}[t] 
 \centering
 \includegraphics[width=0.48\textwidth]{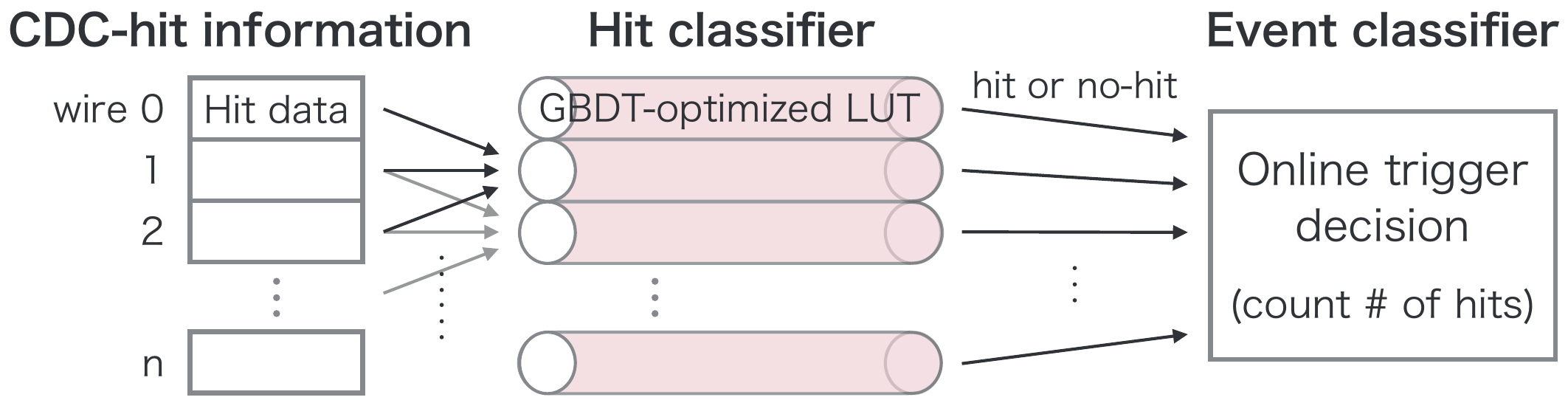}
 \caption[Simple flowchart of the COTTRI system] {Simple flowchart of the COTTRI system.}
 \label{fig:cottri-flow}
\end{figure}

The COTTRI system identifies the conversion-electron events using gradient boosted decision trees (GBDTs), which is one of the ML methods.
A classifier optimized by the GBDTs is implemented in each COTTRI board in the form of lookup tables (LUTs) in FPGA.
The trigger signal flowchart is shown in \figurename~\ref{fig:cottri-flow}.
The LUTs are fed for every adjacent three CDC wires to evaluate hits using the 2-bit energy deposition information, followed by an event classifier which generates the final trigger decision.
%
%

\subsection{Hit Characteristics}
\begin{figure}[t]
 \centering
 \includegraphics[width=0.4\textwidth]{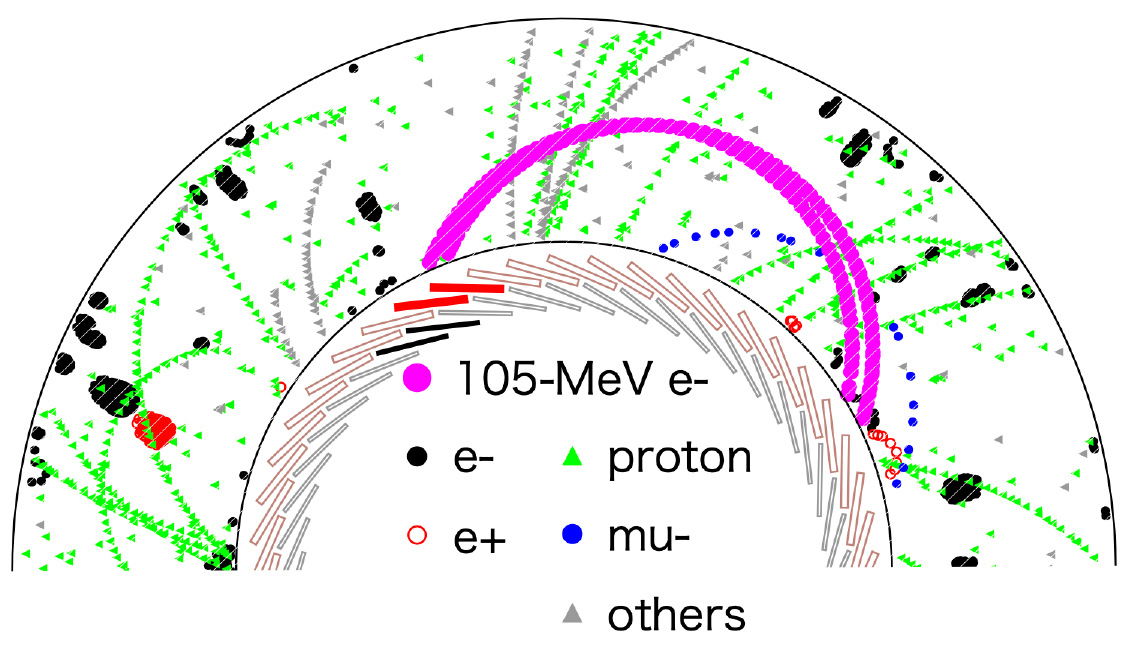}
 \caption[Simulated CDC-hit map including conversion-electron hits] {Simulated CDC-hit map including hits from a 105-MeV conversion electron. Each dot represents the hit position of charged particles. The ``others'' includes heavy particles, such as alpha, triton, and heavy ions. The red and black tilted boxes inside the inner wall of the CDC are Cherenkov counters and scintillators of the CTH, respectively. The filled boxes represent CTH counter hits.  }
 \label{fig:cdc-hit-map}
\end{figure}

In order to design the classification algorithm, it is important to understand the differences of hit and track characteristics between the conversion electron and background particles. 
\figurename~\ref{fig:cdc-hit-map} shows a simulated conversion-electron trajectory overlaying with background particles recorded within an event window of $1.1\,\mathrm{\mu s}$.
%
%
The main background particles are protons from the muon-nuclear-capture processes and low-energy electrons from the gamma-ray interactions at the CDC walls.
Notable differences between background and signal hits appear in the hit patterns and energy deposition.
The conversion electron makes a helical trajectory that is fully contained in the CDC due to the magnetic field, as shown in \figurename~\ref{fig:cdc-hit-map}.
The track will produce a series of neighboring hits in the azimuthal direction at a radius given by the transverse momentum of the conversion electron, and no or very few hits beyond this radius.
The low-energy electrons pass along the CDC wires, and their trajectories are helical orbits with small radii, resulting in long-lived hits on the same wire.
The protons mostly have high momenta and pass through the CDC from inside to outside with a larger energy loss than the conversion electrons.

\subsection{Classification Algorithm}
\begin{figure}[t]
 \centering
 \begin{minipage}{0.4\textwidth}
  \centering
  \includegraphics[width=\textwidth]{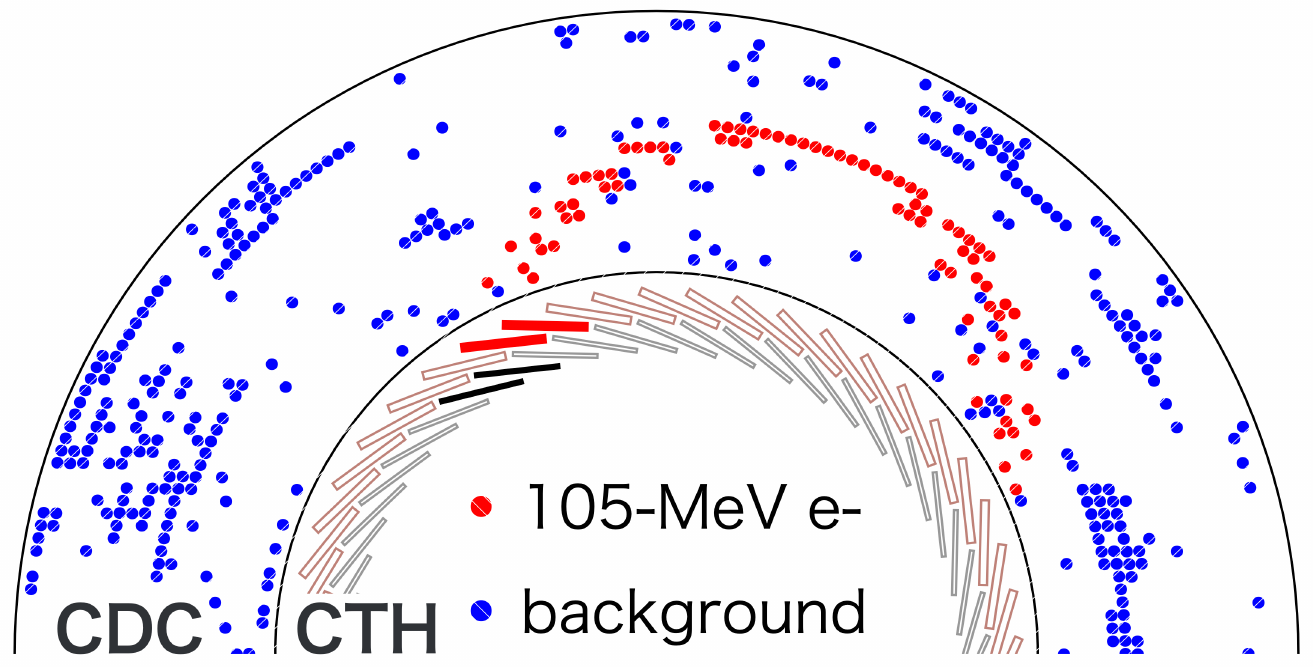} \\ (a)
  \label{fig:cdc-hit-map-before}
 \end{minipage}\\
 \vspace{5pt}
 \begin{minipage}{0.4\textwidth}
  \centering
  \includegraphics[width=\textwidth]{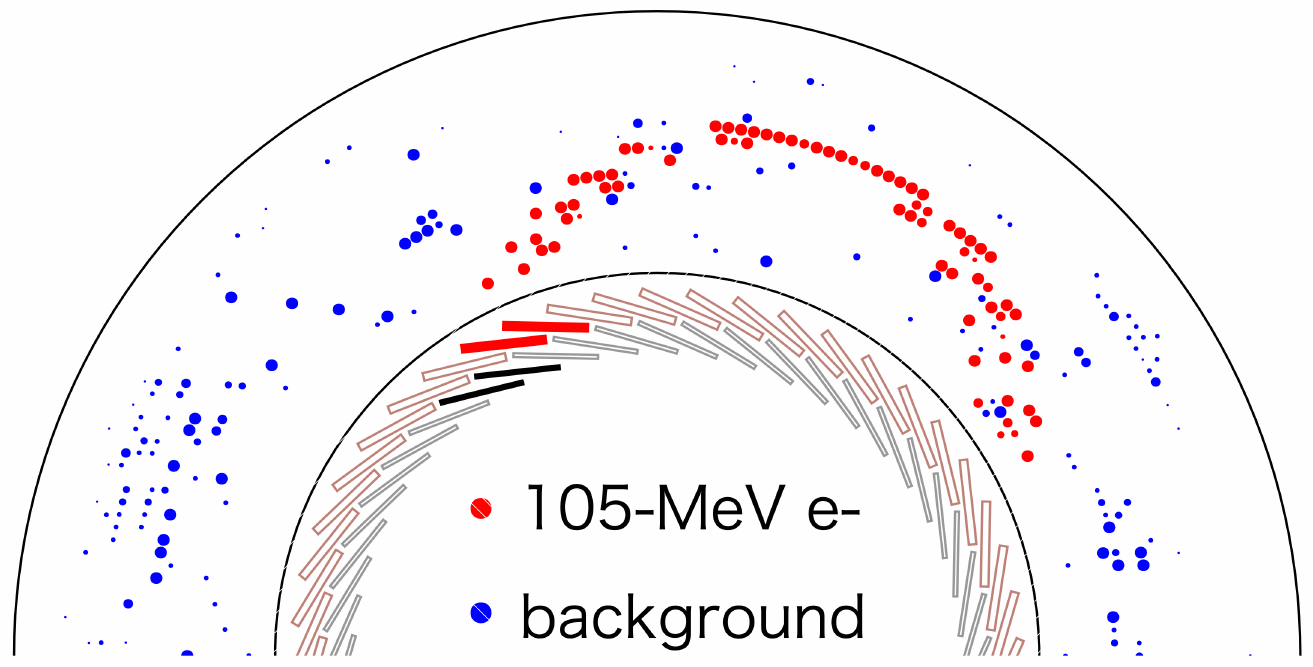} \\ (b)
  \label{fig:cdc-hit-map-after}
 \end{minipage}
\caption[Hit maps of the CDC] {Hit maps of the CDC 
(a) before and (b) after applying the GBDTs. See the text for details.}
 \label{fig:cdc-hit-map-comp}
\end{figure}

In the hit classification stage, GBDTs are used to evaluate whether the hits in the set of neighboring wires are consistent with the expectations for a conversion electron.
The signal-like hits have larger GBDT-output values and are selected for the event classification.
\figurename~\ref{fig:cdc-hit-map-comp} shows the CDC-hit maps before (\figurename~\ref{fig:cdc-hit-map-comp}a) and after (\figurename~\ref{fig:cdc-hit-map-comp}b) applying the GBDTs.
Red and blue dots represent signal and background hits based on simulation information.
The dot size of \figurename~\ref{fig:cdc-hit-map-comp}b reflects the GBDT-output value.
While some background hits with large GBDT-output still remain after applying GBDT, it is clear that GBDT can classify the signal hits out of background hits. 
%
%
Therefore, the deposited energy on the wire of interest and its radial position are selected as the GBDT-input features.
In order to eliminate hits of the low-energy electrons, hit classifiers begin with filtering the wires having long-lived hits.
The energy deposition of neighboring wires in the same layer is also used to suppress low-energy electron hits. 
For the hardware implementation, the input feature must be quantized so that the total size of trigger data fits to the reasonable data transfer rate between different FPGAs with the available FPGA logic resources, such as the number of LUTs.
The energy deposition of each wire is compressed into 2 bits, as written in Section~\ref{sec:cottri}.
Therefore, 6-input LUTs are used for the hit classification using energy deposition from the wire of interest and two neighboring wires. 
We implement a set of 6-input LUTs inside the FPGA, and each set of 6-bit wire hit patterns is fed into each different LUT depending on their radial position.
Thus all the input features (deposited energy, neighboring hit pattern, and radial position) can be considered.

\begin{figure}[t]
 \centering
 \includegraphics[width=0.45\textwidth]{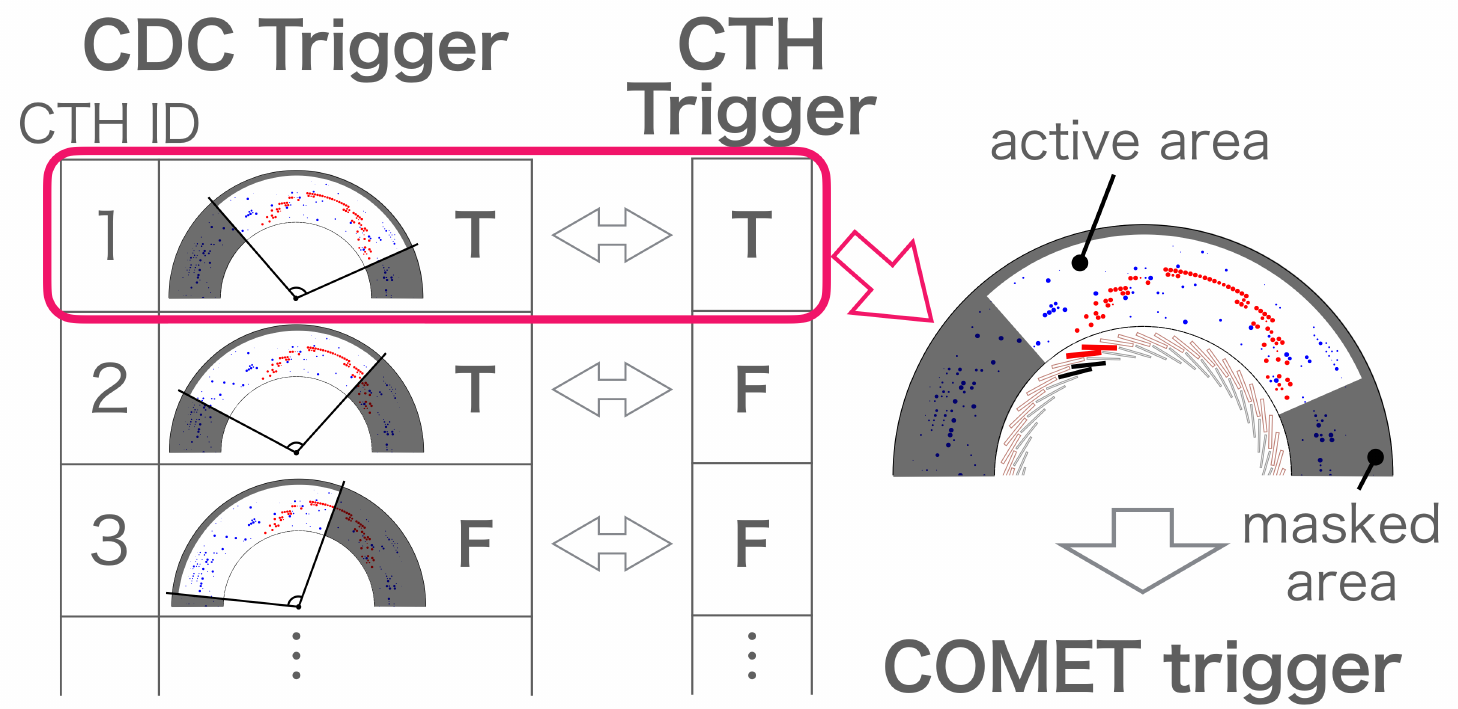}
 \caption[Procedures for the final trigger decision] {Procedures for the final trigger decision. CTH ID means an identification number for each CTH counter. ``T'' (true) and ``F'' (false) mean triggered and non-triggered sections, respectively. Hit counters of the CTH are filled with red for the Cherenkov counters and black for the scintillation counters.}
 \label{fig:cottri-coincidence}
\end{figure}

\figurename~\ref{fig:cottri-coincidence} describes the procedure of the final trigger decision by the event classifier, which combines CDC and CTH trigger information. 
%
%
The conversion electron leaves hits only in a part of the CDC readout area, which is correlated with the CTH-hit positions, as shown in \figurename~\ref{fig:cdc-hit-map}.
An active part of the CDC is defined for each CTH counter to reject background hits efficiently while keeping the conversion-electron hits.
When the number of signal-like hits in each active part exceeds a threshold, the CDC trigger is generated for each CTH counter.
The CTH trigger provides the counter information passing the four-fold coincidence requirement. 
The final trigger decision is performed by taking a coincidence between these CDC and CTH triggers.

\section{Trigger performance evaluation}

\subsection{Timing Considerations}
\begin{figure}[t]
 \centering
 \includegraphics[width=0.35\textwidth]{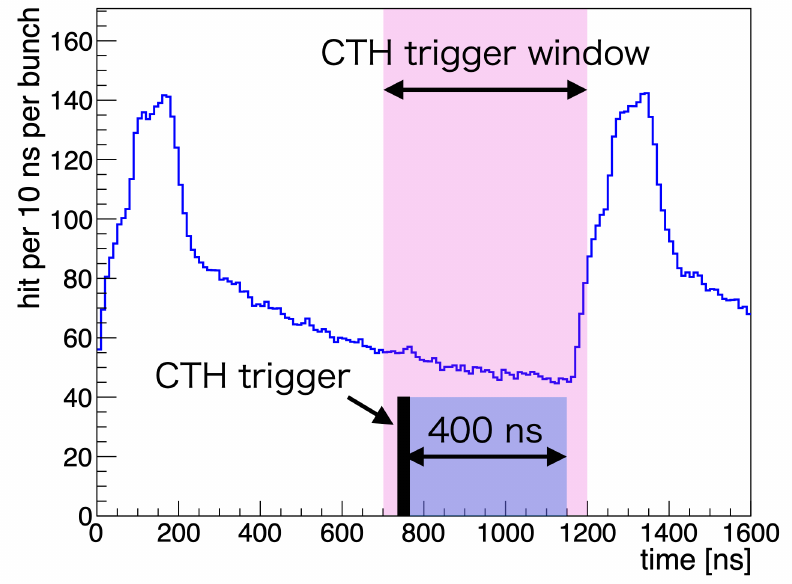}
 \caption[Time distribution of the background hits] {Time distribution of the background hits. The drift time is not considered here. The horizontal axis shows time from a proton bunch. The CTH trigger window is set to [700\,ns, 1200\,ns], and the integration time is set to 400\,ns from the CTH-trigger timing. Detection of the CDC hits is delayed due to the drift time, and the end of the integration time can be extended to 1600\,ns from the proton bunch.}
 \label{fig:param-t}
\end{figure}

For the data taking in the pulsed proton beam, it is important to find an optimal timing window for accepting triggered events to be read out by the DAQ system.
\figurename~\ref{fig:param-t} shows the timing distribution of background particle hits in the CDC.
%
%
After the proton bunches hit the production target, the secondary particles reach the detector with a certain delay in time given by the beam transport system, then the number of background hits immediately increases.
This background level gradually decreases mainly with the lifetime of the stopped muons in the aluminum target.
The bunch separation in the COMET Phase-I experiment is 1170\,ns, and therefore, the background level increases again 1170\,ns with the arrival of the next bunch.
In order to perform the physics measurement during the time of low background, the measurement timing window of the CTH trigger is set to [700\,ns, 1200\,ns] after the proton bunch.

A charged particle passing through the CDC loses energy by ionization.
The RECBE detects signals from the deposited charges in each CDC cell, which arrive within a certain time interval given by the maximal drift-time of the charges to the sense wire in the cell.
Therefore, an integration time is applied for the COTTRI system after the CTH trigger.
In this integration time, the hit classifiers collect the CDC hits and filter wires having long-lived hits.
If the integration time is too short, the number of detected signal hits is small in each trigger-decision time window and genuine conversion electrons may be lost.
On the other hand, if it is too long, many signal hits could be misidentified as long-lived hits and filtered out due to the pile-up criteria, and again, genuine conversion electrons may be lost.
The CDC cell is $16\,\mathrm{mm}\times16.8\,\mathrm{mm}$ in size, and the drift velocity is about $25\,\mathrm{\mu m/ns}$.
The integration time of 400\,ns accepts almost all of the hits.

\subsection{Input Features} \label{subsec:input-features}
\begin{figure}[t]
 \begin{minipage}{0.48\hsize}
  \centering
  \includegraphics[width=\textwidth]{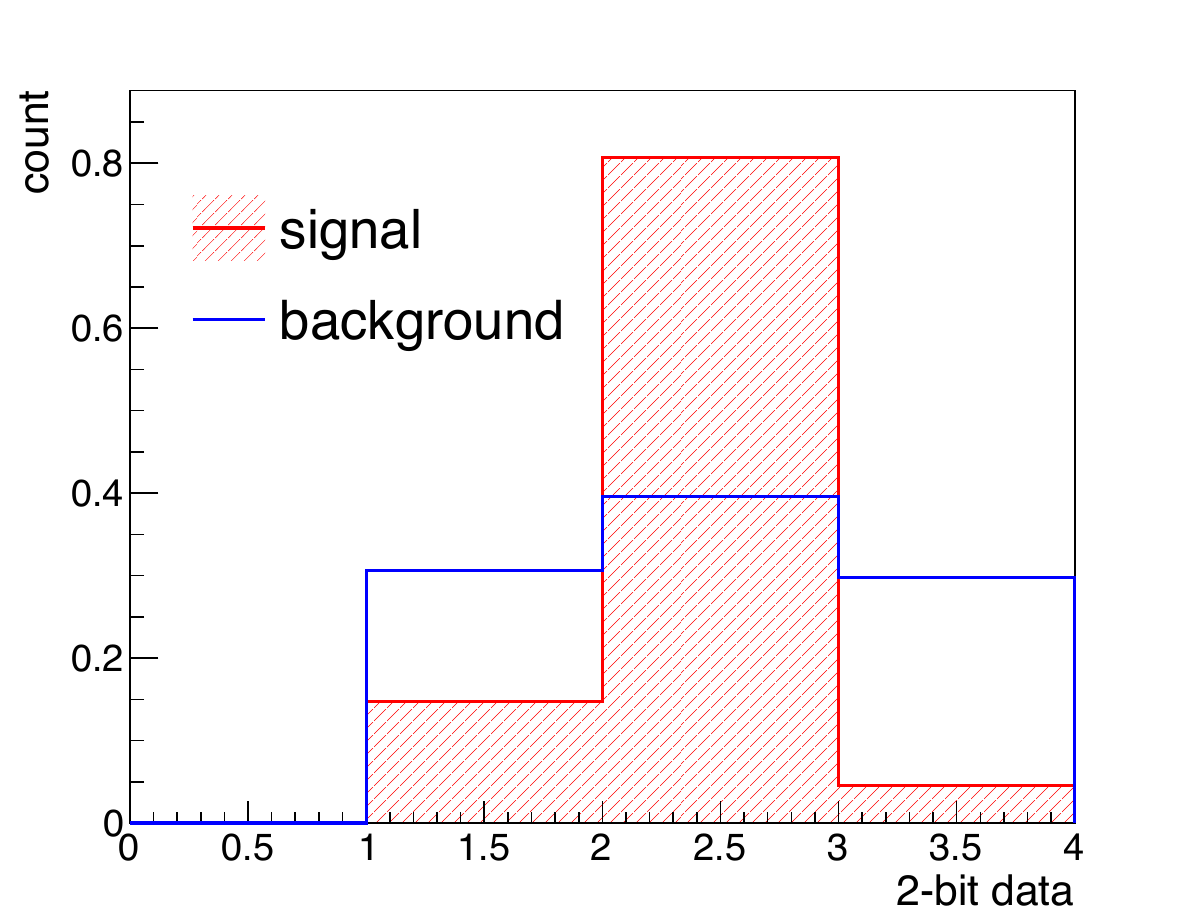}
  (a)
 \end{minipage}
 \begin{minipage}{0.48\hsize}
  \centering
  \includegraphics[width=\textwidth]{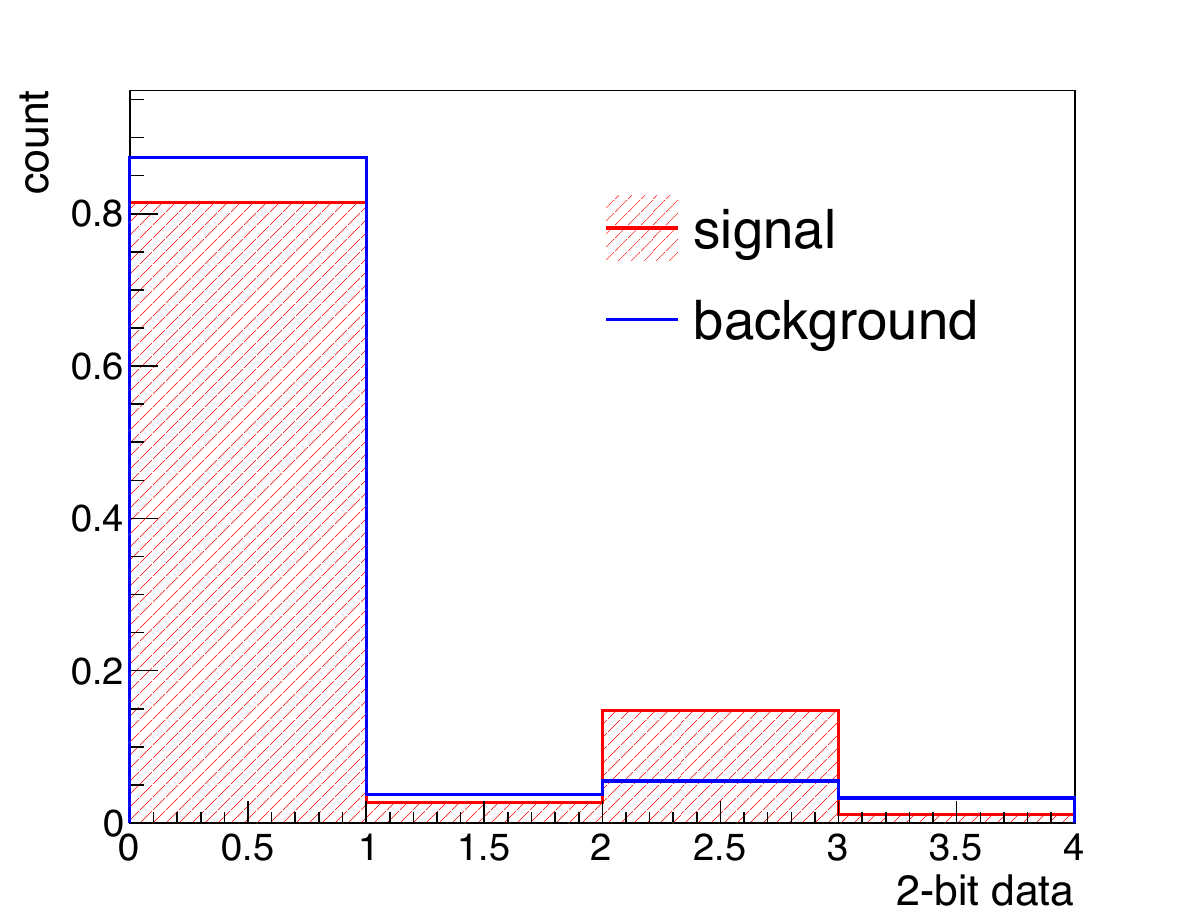}
  (b)
 \end{minipage}
 \caption[Distribution of the 2-bit compressed data] {Distribution of the 2-bit compressed energy deposition data on the wire of interest (a) and the neighboring wires (b).}
 \label{fig:2-bit-data}
\end{figure}

The 2-bit data compression process for the wire energy deposition is optimized to get better signal-event acceptance for the conversion electrons.
\figurename~\ref{fig:2-bit-data} shows the 2-bit energy deposition histogram on the wire of interest and its two neighbors within the same layer, after the optimization.
%
%
Clearer peaks of conversion electron are observed in the 2-bit energy deposition distribution than in the case of background hits, for both the wire of interest and the neighboring wires.

\begin{figure}[t]
 \centering
 \includegraphics[width=0.35\textwidth]{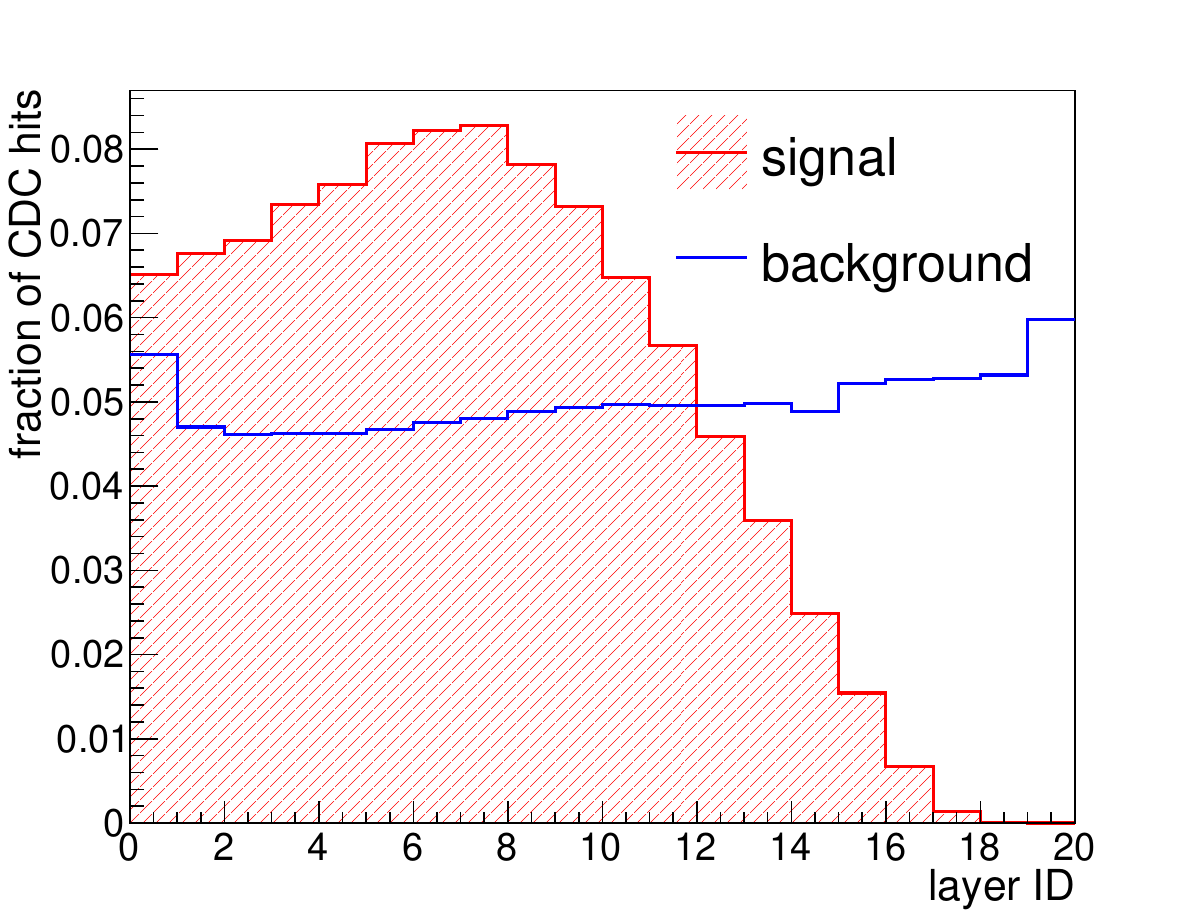}
 \caption[Radial position distribution of the CDC hits] {Radial position distribution of the CDC hits. The layer ID of 0 means the innermost layer of the CDC.}
 \label{fig:cdc-hit-radial-position}
\end{figure}

\figurename~\ref{fig:cdc-hit-radial-position} shows the radial position distribution of CDC hits.
%
%
Since the mono-energetic conversion electron is emitted from the aluminum target located at the center of the CDC, it produces fewer hits in the outer layers of the CDC due to their limited transverse momentum.
In contrast, hits generated by the background particles are distributed more homogeneously in the entire CDC region because of their characteristics, as mentioned in Section~\ref{sec:algo}.
The inner and outer walls of the CDC produce more low-energy electrons.
Because of this, hits in the innermost and the three outermost layers are ignored in the hit classification.

\subsection{Classification Performance}
\begin{figure}[t]
 \centering
 \includegraphics[width=0.35\textwidth]{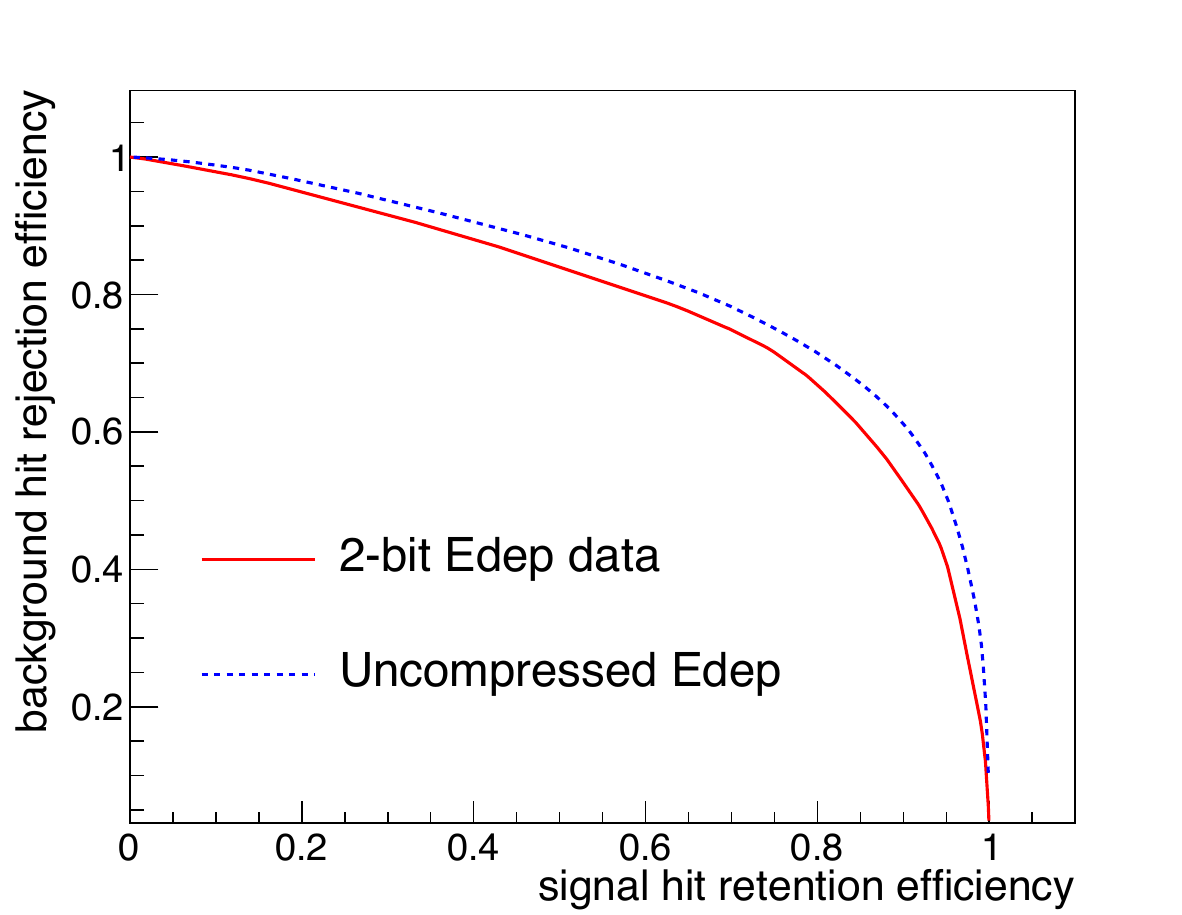}
 \caption[ROC curves of the hit-classification performance] {ROC curves of the hit-classification performance. "Edep" means deposited energy in the CDC.}
 \label{fig:roc-curve-hit-class}
\end{figure}

The GBDT approach in the hit classification was performed by using the Toolkit for Multivariate Data Analysis with ROOT (TMVA), which provides many machine learning techniques for classification and regression~\cite{root-tmva}.
The receiver operating characteristic (ROC) curve of the hit classification is shown in \figurename~\ref{fig:roc-curve-hit-class}.
%
%
The classification result is compared with the uncompressed data case (energy deposition data in 10-bit). 
The 2-bit data case shows worse classification performance than the uncompressed data case, however, it still shows reasonable hit classification performance.
The threshold for the GBDT output is scanned to optimize the event-classification quality.
In this study, this optimization was done by changing the signal-hit retention efficiency from 25\% to 95\% at a 5\% interval.
As a result, it is found that the threshold for the signal-hit retention efficiency of 75\% gives the best performance in the event classification.

\begin{figure}[t]
 \begin{minipage}{0.49\hsize}
  \centering
  \includegraphics[width=\textwidth]{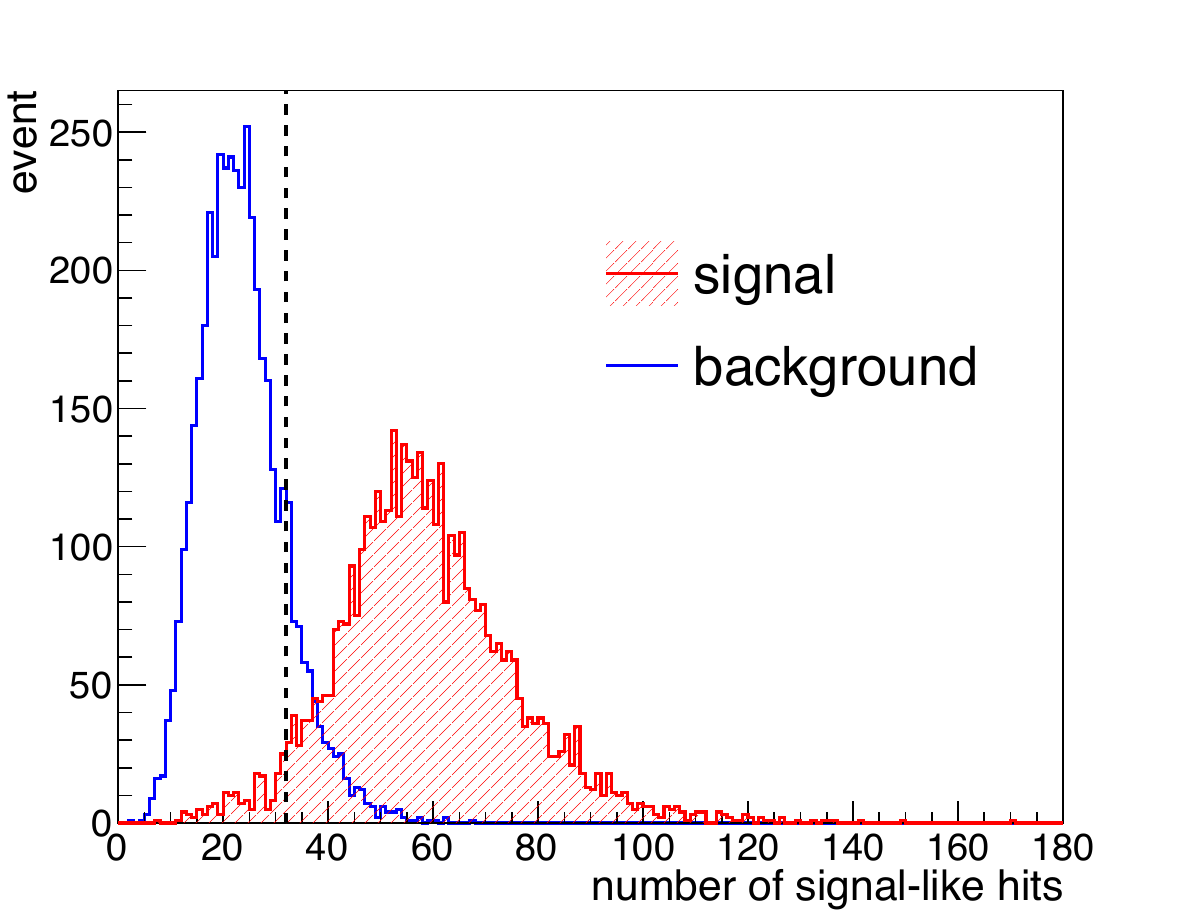}
  \\ (a)
  \label{fig:num-of-hits}
 \end{minipage}
 \begin{minipage}{0.49\hsize}
  \centering
  \includegraphics[width=\textwidth]{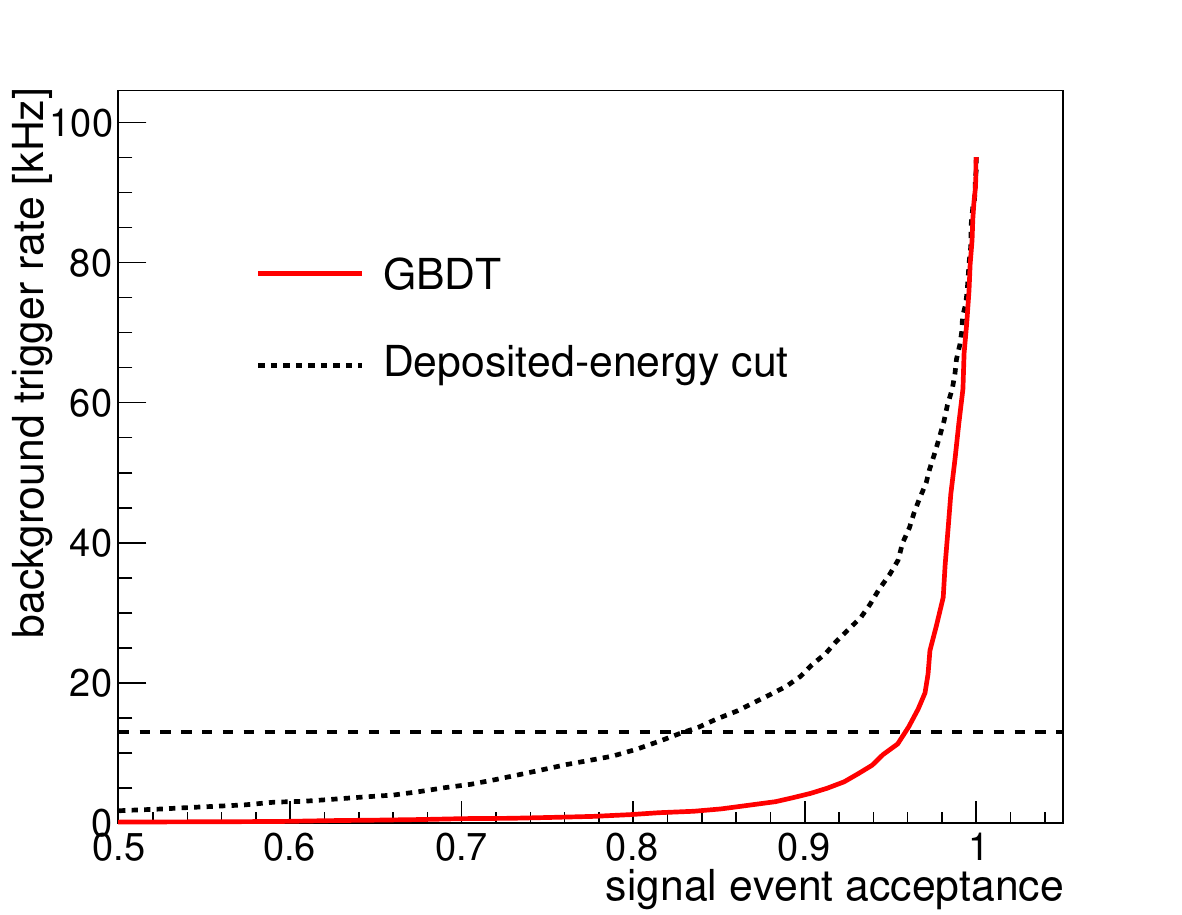}
  \\ (b)
  \label{fig:trigger_rate-vs-acceptance}
  \end{minipage}
  \caption[Number of signal-like hits & background-trigger rate versus signal-event acceptance]{(a) Number of signal-like hits in the active part of the CDC after event classification. The dashed line shows the threshold of 32 which gives the required trigger rate of 13\,kHz. (b) Background trigger rate versus signal event acceptance. The dashed line shows the required trigger rate.}
  \label{fig:event-classification}
\end{figure}

\figurename~\ref{fig:event-classification}a shows the distribution of the number of signal-like hits in both signal and background events after the hit classification.
%
%
Since the signal hits are successfully selected in the hit classification, the signal events contain more signal-like hits than the background events.
From this result, the signal- and background-event acceptance were calculated by applying a threshold for the number of signal-like hits.
The background-trigger rate was estimated by the expected background trigger rate of 91\,kHz, multiplied by the background-event rejection efficiency.
\figurename~\ref{fig:event-classification}b shows the relation between the trigger rate and signal-event acceptance.
According to this, the COTTRI system is expected to provide a signal-event acceptance of 96\% while keeping the required trigger rate of 13\,kHz, which enables the data taking with almost 100\% DAQ efficiency.

Here (see also \figurename~\ref{fig:event-classification}b), this result is also compared with a simpler case, which only applies energy threshold cut to reject higher-energy depositing background hits. 
This method gives a signal-event acceptance of 83\% for the required trigger rate.
It is evident that the GBDT-based classification achieves better performance than the simple energy-threshold cut.

\section{Trigger hardware}
We have designed and produced a new COTTRI FE board and developed the firmware for both the COTTRI FE and MB.
Its operation test was performed to measure the trigger latency and confirm the feasibility of its logic.

\subsection{Hardware Development}
\begin{figure}[t]
 \centering
 \includegraphics[width=0.35\textwidth]{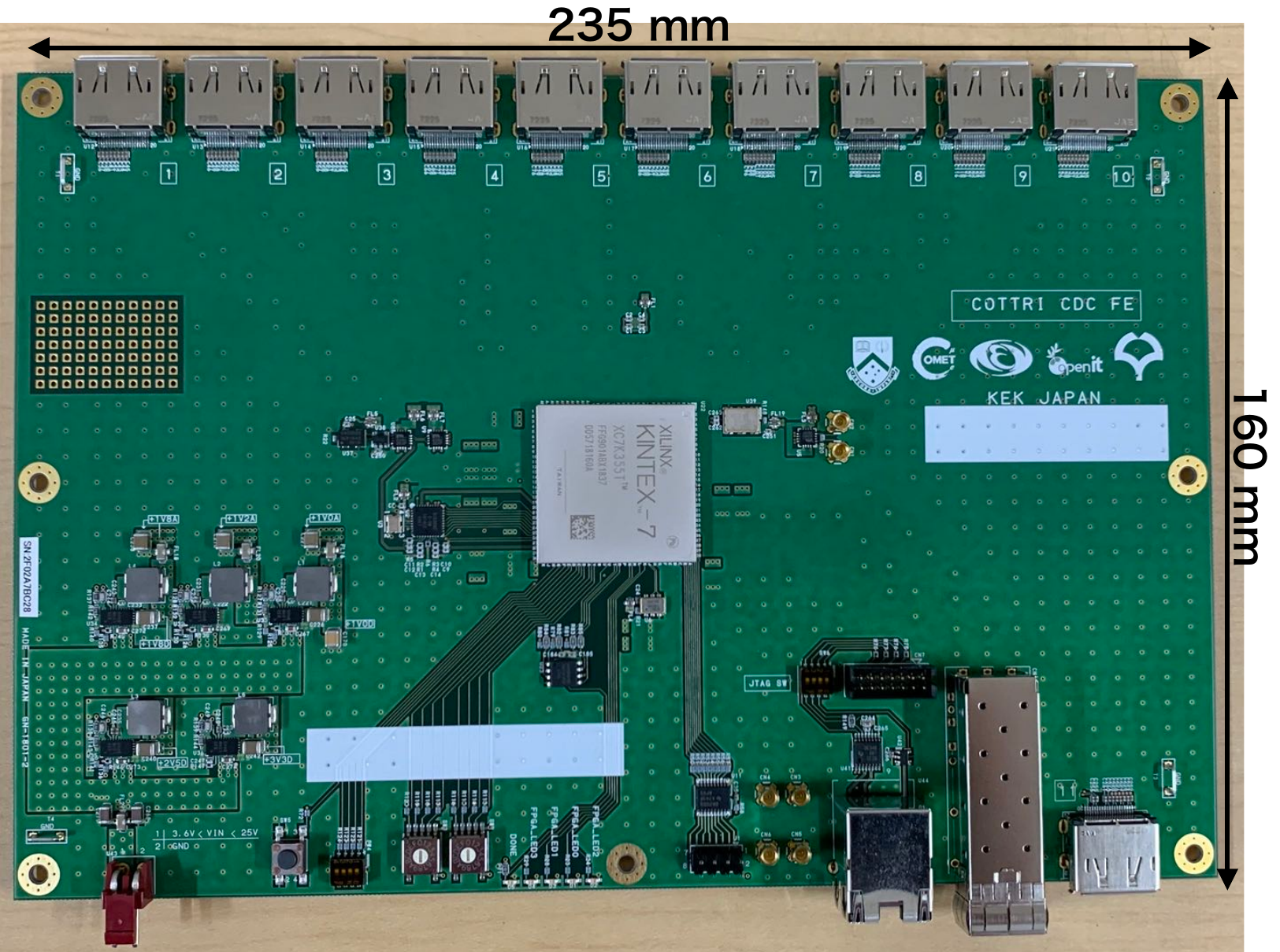}
 \caption[Picture of the COTTRI FE] {Picture of the COTTRI FE.}
 \label{fig:cottri-board}
\end{figure}

The COTTRI FE has been designed and produced, and the functions related to the online trigger algorithm have been implemented on both the COTTRI FE and MB.
\figurename~\ref{fig:cottri-board} shows the COTTRI FE.
%
%
The floor design is similar to the COTTRI MB one which was already developed and tested~\cite{trigger-eps-hep-proc} and contains:
\begin{itemize}
 \item Ten DisplayPorts (DPs) for trigger-related communication,
 \item An FPGA (Kintex7, xc7k355tffg901~\cite{kintex7}) for data processing,
 \item A clock jitter cleaner (Si5326~\cite{si5326}) for generating a clock signal for multi-gigabit transceivers on the FPGA,
 \item An SFP+ port for communication with DAQ system, and
 \item A DP connector for communication with the COTTRI MB.
\end{itemize}

The main roles of the COTTRI FE and MB are the hit classification and the generation of the CDC trigger, respectively.
The COTTRI FEs receive the 2-bit data from the RECBEs corresponding to CDC hit wires and filter out long-lived hit wires in a pre-processing step for the hit classification within the integration time window.
GBDT-optimized LUTs for the hit classification are implemented in the FPGA of the COTTRI FE.
For a 6-input LUT, two built-in reconfigurable 5-input LUTs (CFGLUT5s~\cite{xilinx-libraries}) are implemented on the COTTRI FEs to enable the dynamic adjustment of trigger setup during a run.
These LUTs convert the input features to the GBDT outputs within one clock cycle, and the signal-like hits are selected by using their high LUT outputs as indices.

\begin{figure}[t]
 \centering
 \includegraphics[width=0.25\textwidth]{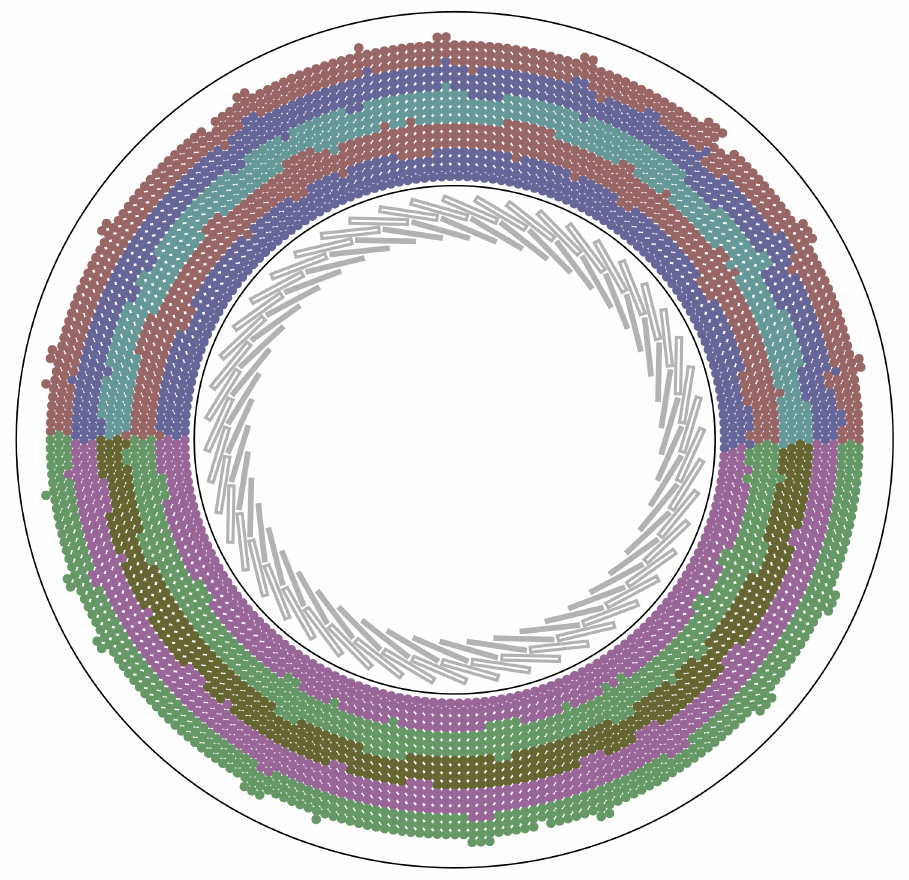}
 \caption[CDC-wire configuration for each COTTRI FE] {CDC-wire configuration for each COTTRI FE. Each color section corresponds to CDC wires grouped by a given COTTRI FE.}
 \label{fig:recbe-cottri-configuration}
\end{figure}

The LUT configuration depends on the CDC wire group for each COTTRI FE.
As described in Section~\ref{subsec:input-features}, the COTTRI system does not require to cover the full readout area of the CDC.
Each COTTRI FE covers eight or nine RECBEs.
\figurename~\ref{fig:recbe-cottri-configuration} illustrates the CDC-wire configurations grouped by the COTTRI FEs.
%
%
Neighboring wires for some wires of interest in the boundary region are not covered by the same COTTRI FE.
For those wires, the dummy data of "2" for neighbors, which look like conversion-electron hits are injected into the LUTs.
After the hit classification, the COTTRI FE counts the number of the signal-like hits for each RECBE.
The COTTRI MB collects those numbers from all the COTTRI FEs, accumulates the total number in each active area, and generates the CDC trigger on the event classification phase.

\subsection{Operation Test}
\begin{figure}[t]
 \centering
 \includegraphics[width=0.4\textwidth]{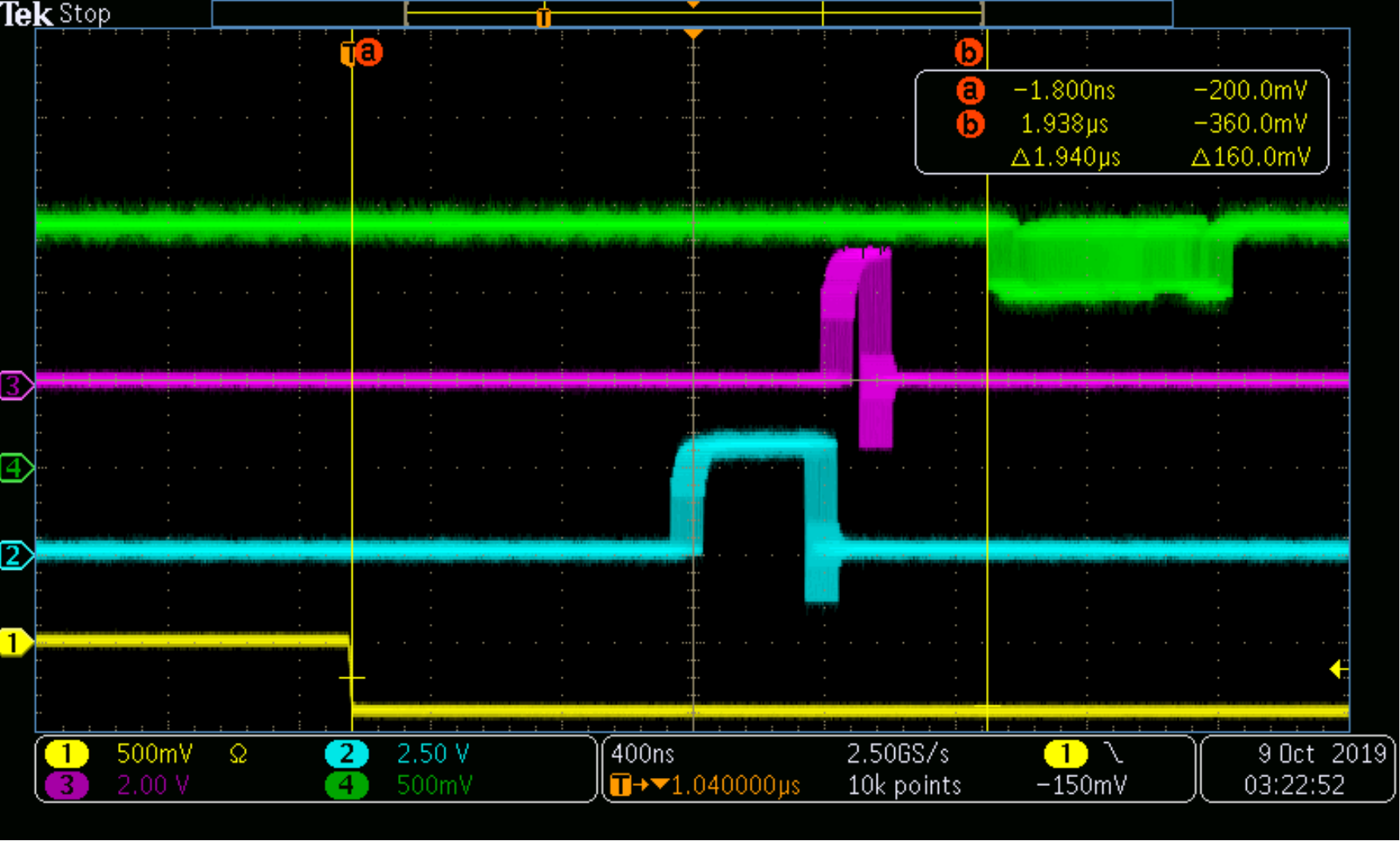}
 \caption[Screenshot of the oscilloscope showing a timing structure of trigger-related signals] {Screenshot of the oscilloscope showing the timing structure of the trigger-related signals. This picture was taken with a persistence display mode. The yellow trace shows the input timing of the test pulse to the RECBE. The light-blue trace shows the timing of finishing the process on the COTTRI FE. The magenta trace shows the timing of asserting the CDC trigger. The green trace shows the timing of the trigger signal received on the RECBE. All traces except for the yellow trace fluctuate within 100\,ns because the RECBE generates and transmits the 2-bit data to the COTTRI system every 100\,ns.}
 \label{fig:trigger-processing-time}
\end{figure}

The COTTRI system was installed in the full chain of the trigger system, and the total trigger latency was measured.
A test pulse from a function generator was utilized as a substitute for a hit signal from the CDC.
An oscilloscope measured the test-pulse timing, trigger latency and jitter.
\figurename~\ref{fig:trigger-processing-time} is a screenshot of the oscilloscope, which shows the timing structure of trigger-related signals.
%
%
Since the 2-bit data is generated and transmitted to the COTTRI system every 100\,ns, the measured time width fluctuates within 100\,ns.
The processing time in the full trigger system was measured to be 2.0$\,\mathrm{\mu s}$ at most, and an integration time of 0.4$\,\mathrm{\mu s}$ must be added.
In addition, some delay for receiving the trigger at the RECBE has to be added because the RECBE recognizes the trigger at the last bit of the trigger number sent from the central system.
The length of this trigger number is 32\,bit, and it is received in synchronization with a 40-MHz clock; the trigger receiving time is 0.8$\,\mathrm{\mu s}$.
Then, the total trigger latency is estimated to be 3.2$\,\mathrm{\mu s}$ at most.
It meets the requirement of less than 7$\,\mathrm{\mu s}$.
These values are summarized in Table~\ref{tb:trigger-latency-table}.
%
\begin{table}[t]
 \begin{center}
 \caption[Processing time in the trigger system]{Processing time in the trigger system.}
 \begin{tabular}{lc}
    \hline
    Processing time in the full trigger system  & $1.9\,\mathrm{\mu s}$ to $2.0\,\mathrm{\mu s}$\\
    Integration time for the drift time         & $0.4\,\mathrm{\mu s}$ \\
    Trigger receiving time on the RECBE         & $0.8\,\mathrm{\mu s}$\\\hline
    Total       & $3.1\,\mathrm{\mu s}$ to $3.2\,\mathrm{\mu s}$ \\
    \hline
\end{tabular}
 \label{tb:trigger-latency-table}
 \end{center}
\end{table}
%

\begin{figure}[t]
 \centering
 \includegraphics[width=0.4\textwidth]{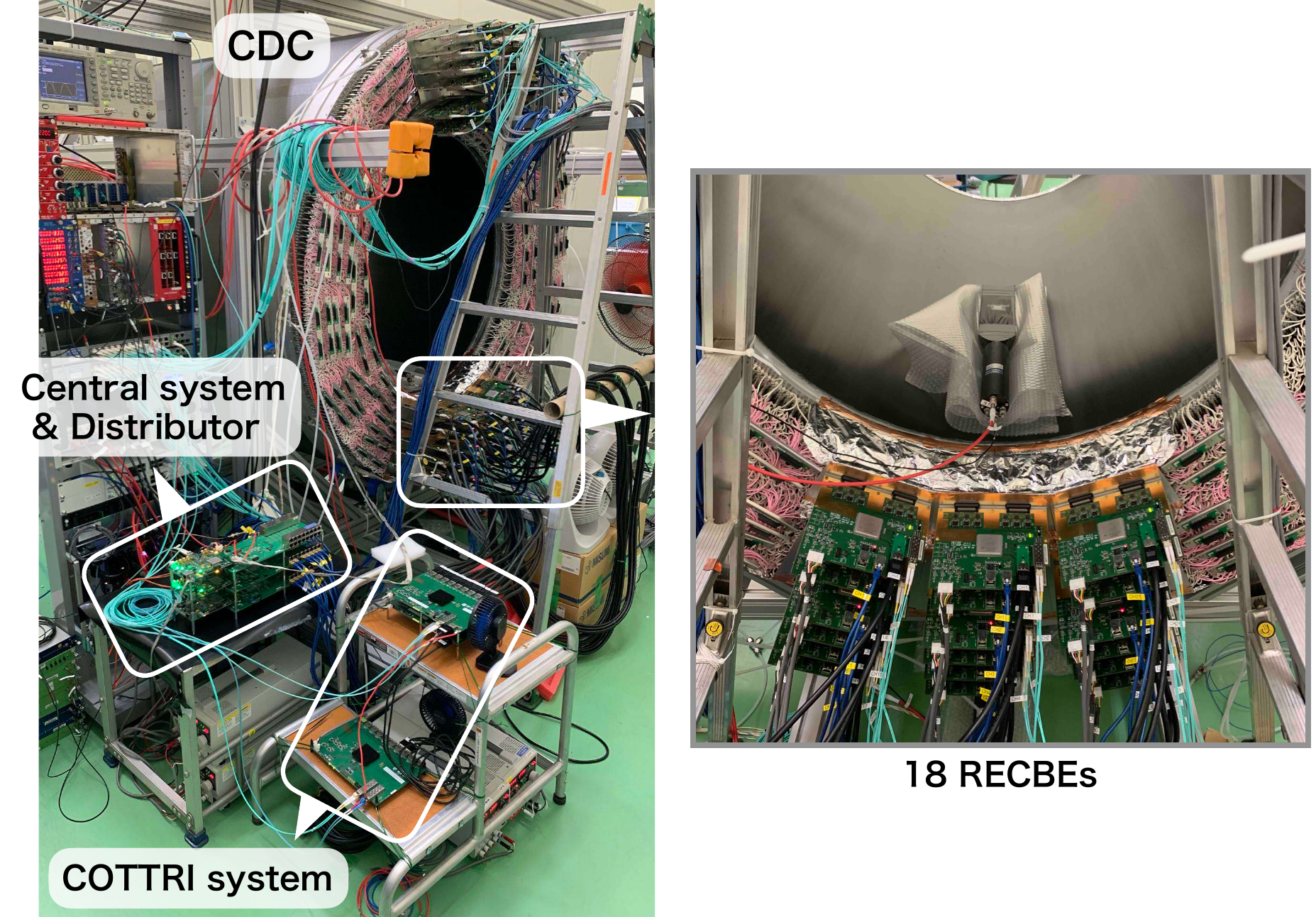}
 \caption[Photographs of the operation test for the COTTRI system]{Photographs of the operation test for the COTTRI system.}
 \label{fig:cdc-cosmic-ray-test}
\end{figure}

\begin{figure}[t]
 \centering
 \includegraphics[width=0.3\textwidth]{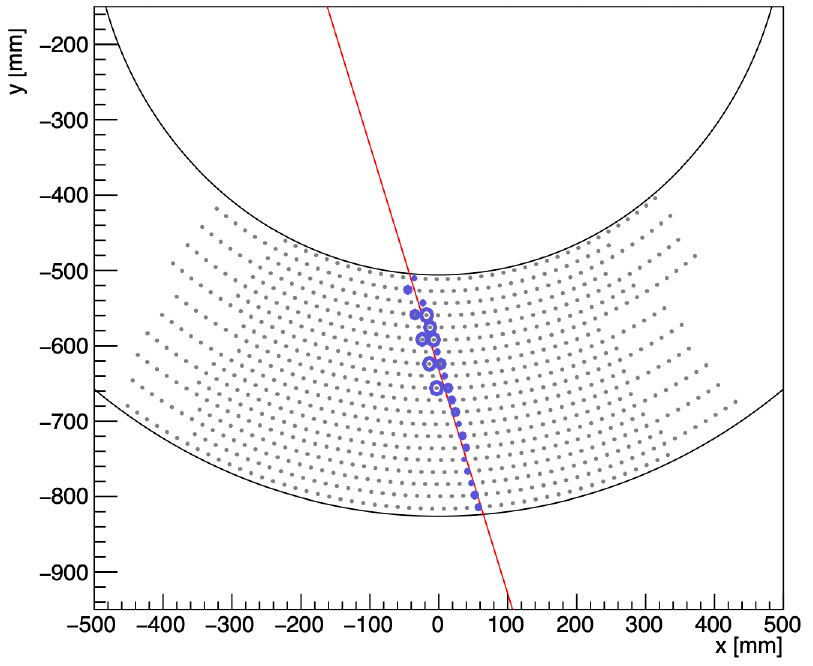}
 \caption[Display of a cosmic-ray event]{Display of a reconstructed cosmic-ray event. The gray dots represent the CDC wires. The red line represents the reconstructed cosmic-ray track. The blue circles represent the hit position and the drift circles whose radii correspond to the drift distance calculated from the drift time in each cell.}
 \label{fig:cdc-cosmic-ray-event}
\end{figure}

In order to confirm that the COTTRI system works correctly, an operation test was performed in a cosmic-ray experiment using 18 RECBE boards.
The test setup is shown in \figurename~\ref{fig:cdc-cosmic-ray-test}.
%
%
The trigger decision was made by counting the number of CDC hits in the COTTRI system without any other trigger signal, such as the CTH.
The COTTRI system successfully detected and triggered cosmic rays by applying the CDC self-trigger, as shown in \figurename~\ref{fig:cdc-cosmic-ray-event}.
%
%

\section{Summary}
The COMET Phase-I experiment searches for the \muec.
The use of the highly intense muon beam of J-PARC will allow us to reach a 100 times better experimental sensitivity for New Physics searches, but also would lead to an unacceptably high trigger rate for the DAQ system.
For stable and highly efficient data acquisition, we have developed an FPGA-based online trigger using a GBDT-optimized classifier.
Its performance was estimated in simulation.
As a result, this trigger system is expected to provide a signal-event acceptance of 96\% while suppressing the trigger rate from 91\,kHz to the required rate of 13\,kHz.
We have developed the trigger-related electronics and implemented the trigger algorithm.
The latency measurement and the operation test using a part of the CDC readout region were carried out successfully.
The total trigger latency was measured to be 3.2$\,\mathrm{\mu s}$.
The estimated performance of the COTTRI system satisfies the requirements for the trigger rate and the trigger latency.
We conclude that the COTTRI system will solve the trigger rate issue in the COMET Phase-I experiment.

\end{document}